\documentclass[10pt,twocolumn,letterpaper]{article}

\usepackage{cvpr}
\usepackage{times}
\usepackage{epsfig}
\usepackage{graphicx}
\usepackage{amsmath}
\usepackage{amssymb}
\usepackage{caption}
\cvprfinalcopy 

\begin{document}

\title{Convolutional Video Steganography with Temporal Residual Modeling}

\author{Xinyu Weng$^{1,2}$, Yongzhi Li$^{1,2}$, Lu Chi$^{1}$, Yadong Mu$^{1,2}$\\
$^{1}$Institute of Computer Science \& Technology\\
$^{2}$Big Data Scientific Research Center\\
Peking University, China\\
{\tt\small wengxy@pku.edu.cn, yongzhili.pku@gmail.com, chilu@pku.edu.cn, myd@pku.edu.cn}
}

\maketitle

\begin{abstract}
\let\thefootnote\relax\footnotetext{The first two authors contribute equally. Yadong Mu is the corresponding author of this work.}

Steganography represents the art of unobtrusively concealing a secrete message within some cover data. The key scope of this work is about visual steganography techniques that hide a full-sized color image / video within another. A majority of existing works are devoted to the image case, where both secret and cover data are images. We empirically validate that image steganography model does not naturally extend to the video case (\emph{i.e.}, hiding a video into another video), mainly because it completely ignores the temporal redundancy within consecutive video frames. Our work proposes a novel solution to the problem of video steganography. The technical contributions are two-fold: first, the residual between two consecutive frames tends to zero at most pixels. Hiding such highly-sparse data is significantly easier than hiding the original frames. Motivated by this fact, we propose to explicitly consider inter-frame residuals rather than blindly applying image steganography model on every video frame. Specifically, our model contains two branches, one of which is specially designed for hiding inter-frame difference into a cover video frame and the other instead hides the original secret frame. A simple thresholding method determines which branch a secret video frame shall choose. When revealing the concealed secret video, two decoders are devised, revealing difference or frame respectively. Second, we develop the model based on deep convolutional neural networks, which is the first of its kind in the literature of video steganography. In experiments, comprehensive evaluations are conducted to compare our model with both classic least significant bit (LSB) method and pure image steganography models. All results strongly suggest that the proposed model enjoys advantages over previous methods. We also carefully investigate key factors in the success of our deep video steganography model.
\end{abstract}

\section{Introduction}

\begin{figure}[t!]
\begin{center}
\includegraphics[width=1\linewidth]{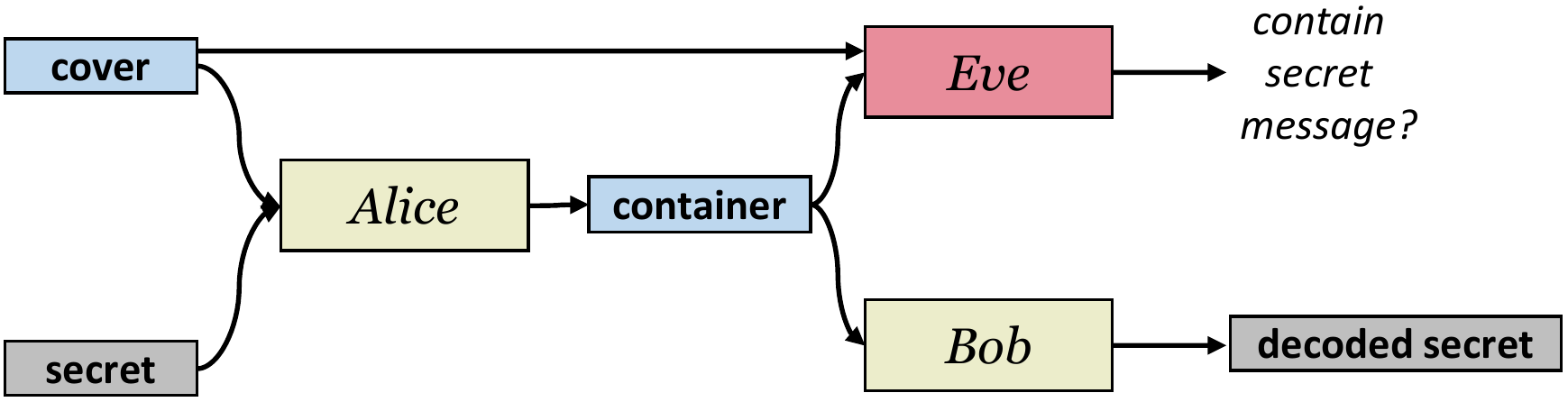}
\end{center}
   \caption{\small The full scheme of steganography. See main text for more explanation.}
\label{fig:stego}
\end{figure}

The term steganography~\cite{hayes17,Baluja17} can date back to some ancient technique developed in the 15th century. The goal of steganography is to encode a secret message in some transport medium (called \emph{cover} in this paper) and covertly communicate with a potential receiver who knows the decoding protocol. Essentially different from cryptography, steganography aims to hide the presence of secret communications, allowing only the target recipient to know. State differently, the covering medium can be publicly visible and yet only the target receiver can perceive the presence and decode the secret message. In practice, any steganography model should conceal a secret message by concurrently optimizing two criteria: minimizing the change of the covering medium that leads to suspect from an adversary, and reducing the residual between decoded secret message and its ground truth. The research on steganography has practical implications. For example, a number of nefarious applications of steganography techniques are known, such as hiding commands that coordinate criminal activities through images posted on social media websites. In the industry of digital publishing, a common tactic to claiming authorship without compromising the integrity of the digital content is to embed digital watermarks. For some brief introduction to steganography, one can refer to~\cite{MorkelEO05,AbboudMY10,KesslerH11,DBLP:journals/tifs/LiTWH14}.

\begin{figure*}[t!]
\begin{center}
\includegraphics[width=0.8\linewidth]{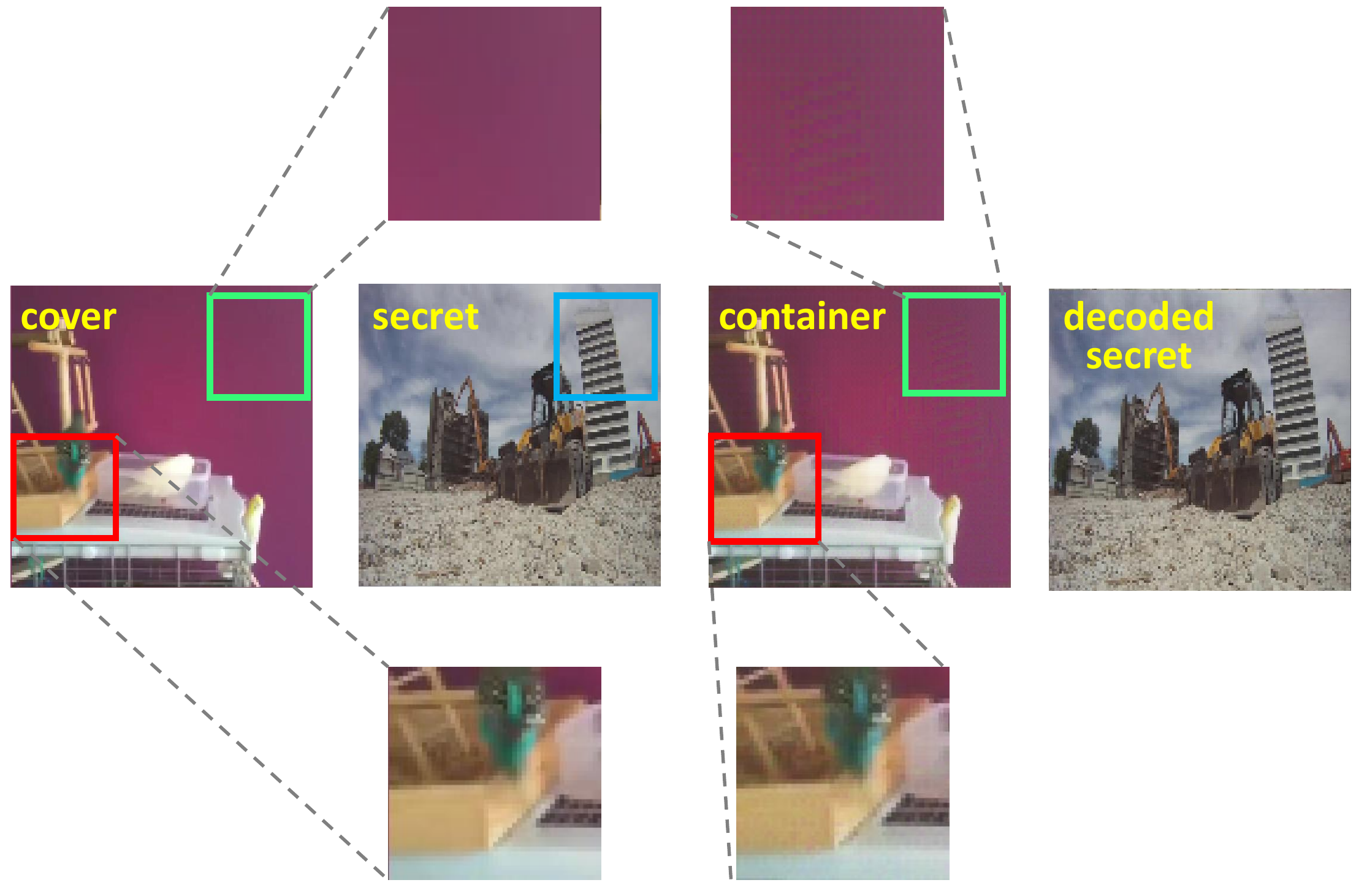}
\end{center}
   \vspace*{-0.1in}
   \caption{\small Exemplar results generated by an image steganography model. The role of each image is depicted in bold yellow text located in the top-left of each image. To depict how container image deviates from the original cover image, we choose two local patches and contrast them for these two images. Indeed, for the local patch delimited by the green box, from the container image one can observe the ghost image of specific building in the secret image (in blue box). Better viewing after enlarging.}
\label{fig:pp_image}
\vspace{-0.1in}
\end{figure*}

Let us first explain the process of a typical steganography system, which is shown in Fig.~\ref{fig:stego}. In classic steganography, the process involves three parties: Alice, Bob and Eve. Alice first conceals a \emph{secret} message into a \emph{cover} to obtain a \emph{container} message (or \emph{steganographic} message), and then sends the container message to Bob. Eve is an adversary (the \emph{steganalyzer}) to both Alice and Bob. Each message that Eve observes is either cover or container. And Eve makes a binary classification on each message. His goal is to judge whether a message is steganographic or not. However, this steganalyzer is not requested to decode the hidden secret message. In this scheme, we say Alice performs perfectly if she ensures: 1) Bob receives the container message and successfully recover secret message at high accuracy using a decoding protocol; and 2) Eve, who always attempts to detect the presence of secret message, has exactly 50\% chance of correctly judging a container or cover message. It is similar to the expectation in adversarial training~\cite{GoodfellowPMXWOCB14,abs-1802-08195}. To accomplish both goals, the container message should not deviate from the original cover too much, avoiding that abnormal pattern appears and is detected by Eve. Meanwhile, it should also be in a good shape to be accurately deciphered by the decoder model at Bob's hand.

Hiding messages in an image has been a long-standing research task of salient practical interest. One can gauge the amount of concealed information through bits-per-pixel (bpp), namely the amortized bits hidden at each pixel in the cover image. A recent research trend is hiding a full-sized color image into another same-sized image as exemplified in~\cite{Baluja17}. We hereafter term the task image steganography. This represents a highly challenging task since it pursues a bpp level of 1 (\emph{i.e.}, each pixel in the cover hides a complete RGB color). Fig.~\ref{fig:pp_image} illustrates a group of typical results calculated from an image steganography model, faithfully following the scheme depicted in Fig.~\ref{fig:stego}. The steganography model can hardly accomplish both of Alice's two goals in the container. As shown in Fig.~\ref{fig:pp_image}, artifacts can often be observed in container, making it easily detected by an adversary. In recent year, to improve the performance of an image stegaongraphy model, researchers have explored deep neural networks for learning both encoding model (\emph{cover} + \emph{secret} $\rightarrow$ \emph{container}) and decoding model (\emph{container} $\rightarrow$ \emph{decoded secret}) in above process. Successful applications are found in~\cite{hayes17,Baluja17}.

\begin{figure*}[t]
\begin{center}
\includegraphics[width=0.95\linewidth]{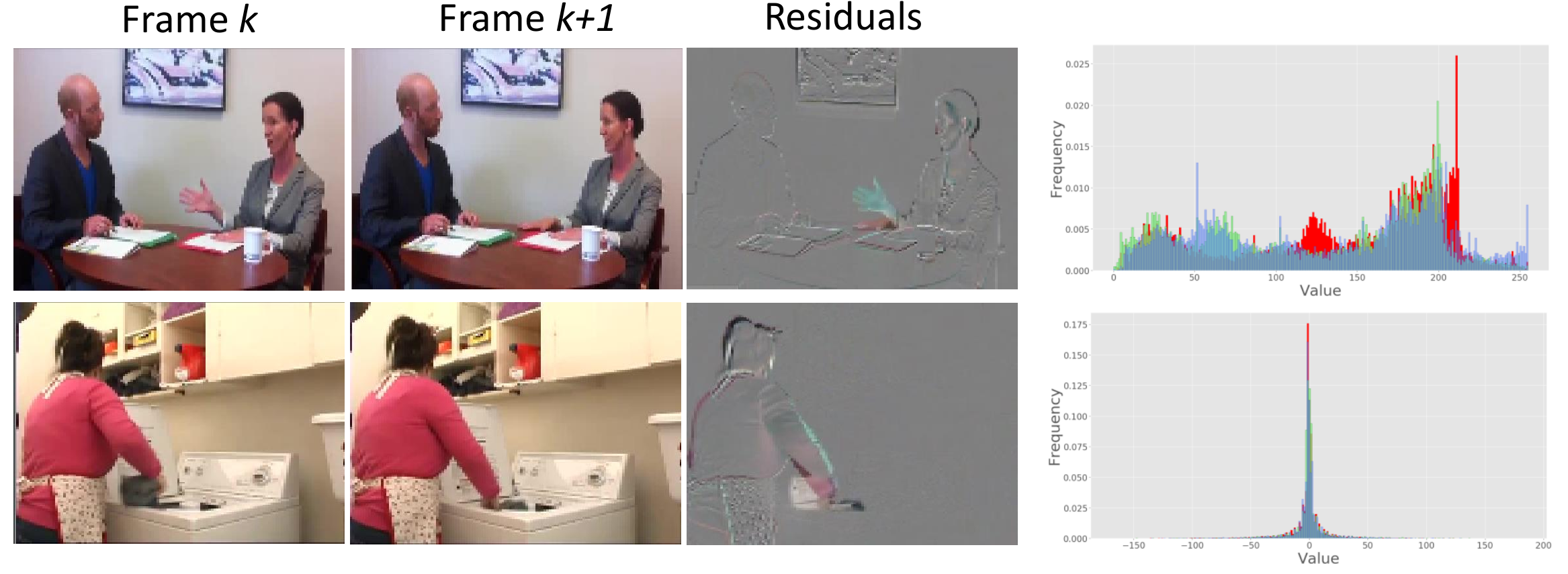}
\end{center}
   \vspace*{-0.1in}
   \caption{\small Examples of video frames and inter-frame residuals. The column \emph{residuals} represent the per-pixel difference between frame $k$ and $k+1$. The righmost column shows the distribution of RGB values (top) and residual values (bottom) for the first frame pair (top row).}
\label{fig:residual_example}
\vspace{-0.1in}
\end{figure*}

In this work, our major focus is video steganography. The task aims to hide a full-sized video clip into another. Considering the increasing popularity of video data across the Internet, the research of video steganography, though currently rarely found in the literature, represents a nascent research topic of key practical implications. One may argue that image steganography model can be readily used to solve the video steganography problem, by pairing frames in cover / secret videos and feeding them into an image model. We argue that this tactic is not optimal, because it does not fully consider the temporal redundancy within consecutive video frames. Our work proposes a novel solution to video steganography. Briefly speaking, the technical contributions are two-fold:

First, the residuals between two consecutive frames are highly sparse. Critically, compared with hiding frame into another frame, hiding such sparse residual in another video frame defines a much easier task. Motivated by this fact, instead of blindly applying image steganography model on all frames, we propose to split frames into two sub-sets: \emph{reference frames} and \emph{residual frames}. Each residual frame is obtained by differencing with specific reference frame. Correspondingly, our model contains two branches at both the encoding and decoding stages, tackling either type of frames respectively. We empirically validate this treatment can significantly boost the container's perceptual quality and increase the possibility of fooling an adversary.

Secondly, our model is fully based on deep convolutional neural networks, which is the first of its kind in video steganography. Specifically, our deep video steganography model consists of two H-networks for hiding references or residuals, and two R-networks for revealing the secret video. The full model is trained without any human annotations and network parameters are optimized from scrach. In experiments, comprehensive evaluations are conducted to validate the powerful modeling of deep networks. We also carefully design ablation investigation to find key factors in our deep video steganography model.

The remainder of this paper is organized as following: We first briefly review the related work in Section~\ref{sec:related}. Section~\ref{sec:model} details the proposed two-branch deep neural networks for the video steganography task. All experimental evaluations and in-depth analysis are found in Section~\ref{sec:exp}. Finally, Section~\ref{sec:conclusion} concludes this work and points out several future research directions.

\section{Related Work}
\label{sec:related}

\begin{figure*}[t]
\begin{center}
   \includegraphics[width=0.7\linewidth]{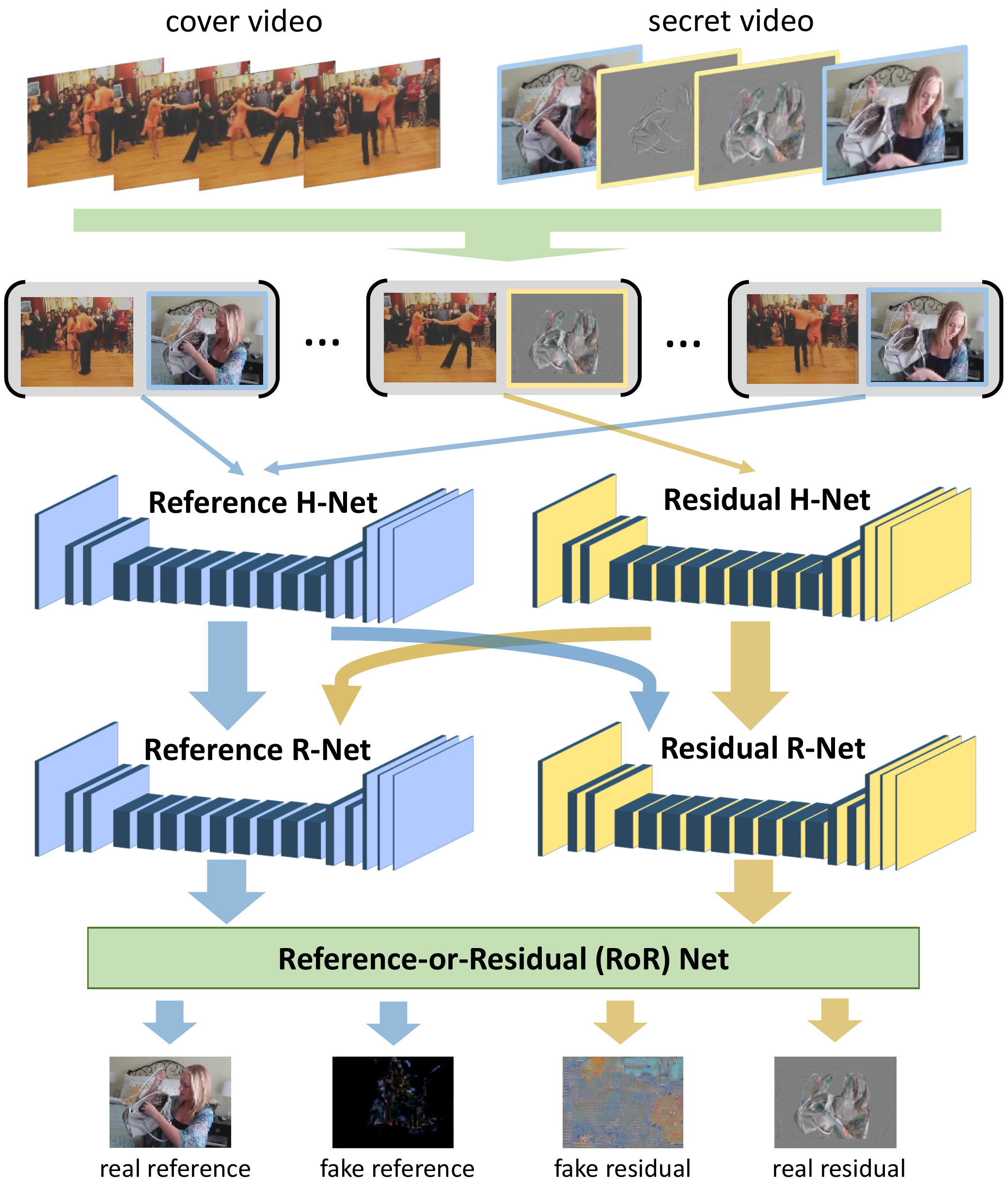}
\end{center}
   \vspace*{-0.1in}
   \caption{\small The computational pipeline of our proposed video steganography model. See text for more details.}
\label{fig:pipeline}
\vspace{-0.1in}
\end{figure*}

Least significant bit (LSB)~\cite{Mielikainen06a,DBLP:journals/corr/abs-1709-06727} is a classic steganographic algorithm. It adopts a simple idea to embed a secret image into another. In digital images, each pixel in an image is comprised of three bytes (\emph{i.e.}, 8 binary bits), representing the RGB chromatic values respectively. The LSB algorithm replaces the least 4 significant bits of the cover image by 4 most significant bits of the secret image. For each byte, the significant bits dominate the color values. This way, the chromatic variation of the container image (altered cover) is minimized. Decoding the concealed secret image can be simply accomplished by reading the 4 least significant bits and performing bit shift. Despite that its distortion is not often visually observable, LSB is unfortunately highly vulnerable to steganalysis~\cite{DBLP:journals/ieeemm/FridrichGD01} - statistical analysis can easily detect the pattern of altered pixels. Recent works have been devoted to more sophisticated methods that preserve the image statistics or design special distortion functions, such as HUGO~\cite{PevnyFB10}, WOW (wavelet obtained weights)~\cite{HolubF12}, S-UNIWARD~\cite{HolubFD14}, and ATS~\cite{Lerch-HostalotM16}.

The most relevant works to ours are two deep learning based image steganography methods in~\cite{hayes17,Baluja17}. Much earlier works~\cite{DBLP:journals/nca/HusienB15,DBLP:journals/corr/PibreJIC15} adopts deep neural networks to elevate accuracies, yet mostly in the decoding process, such as determining which bits to extract from the container images. Both of~\cite{hayes17,Baluja17} build the whole system based on deep networks, including encoding (hiding), decoding (revealing) and adversarial networks. The quantitative evaluations strongly corroborate the superior modeling ability of deep networks. However, to our best knowledge, there is no prior work that explore deep networks for the hiding-video-in-video setting. This paper provides clear evidence that direct adaptation of image steganography model to video data is not an optimal choice and we are thus motivated to devise special video steganography model based on temporal residual modeling.

\section{The Proposed Model}
\label{sec:model}

Fig.~\ref{fig:residual_example} illustrates some motivating fact to our video steganography model. As seen, the residual values between consecutive video frames are dominated by near-zero values. Hiding such high-sparse data into a cover frame intuitively requires less effort compared with a full-colored secret frame, since hiding a zero value is trivial. This way, the cover image tends to be less altered, which potentially increases the chance of fooling an adversary. Using residuals as the secrete message instead can ease Alice's job (or the encoding model) in Fig.~\ref{fig:stego} and meanwhile does not make Bob's task harder. However, to operate on residuals, there are two challenges that we should concern: how to determine encoding the original video frame or its residual with respect to the previous frame? And at the decoding stage, how the decoder knows the received image conceals a full-colored frame or a residual array?

To address above issues, we categorize all secret frames to be either \emph{reference frame} or \emph{residual frame}. Correspondingly, we propose to use two separate encoding / decoding networks for tacking different type of frames. The architecture of our proposed system is shown in Fig.~\ref{fig:pipeline}. The system is comprised of five computational steps:

\begin{table*}[thb]
\centering
\caption{Architecture of both Reference Hiding network and Residual Hiding network. There is a batch normalization layer(BN) and a Leaky Rectified Linear Unit(LeakyReLU)  after each convolution layer. And there is  a BN and a Rectified Linear Unit(ReLU) after each deconvolution layer except the last one. The output deconvolution layer is followed by a $Sigmoid$ function.}\label{tab:h}
\begin{tabular}{cccccccc}
\hline
Index & Type& Kernel& Stride&Padding& Input & Out & Concatenation\\
\hline
1 & Conv2d.& 4$\times$4& 2& 1& 6 & 64 & N/A\\
2 & Conv2d.& 4$\times$4& 2& 1& 64 & 128 & N/A\\
3 & Conv2d.& 4$\times$4& 2& 1&128& 256 & N/A\\
4 & Conv2d.& 4$\times$4& 2& 1& 256 & 512 & N/A\\
5 & Conv2d.& 4$\times$4& 2& 1& 512 & 512 & N/A\\
6& Conv2d.& 4$\times$4& 2& 1& 512 & 512 & N/A\\
7& Conv2d.& 4$\times$4& 2& 1& 512 & 512 & N/A\\
\hline
8 & deConv2d.& 4$\times$4& 2& 1& 512 & 512 & N/A\\
9 & deConv2d.& 4$\times$4& 2& 1& 1024 & 512 &concat with layer \#6\\
10& deConv2d.& 4$\times$4& 2& 1& 1024 & 512 &concat with layer \#5\\
11& deConv2d.& 4$\times$4& 2& 1& 1024 & 256 &concat with layer \#4\\
12 & deConv2d.& 4$\times$4& 2& 1& 512 & 128 &concat with layer \#3\\
13 & deConv2d.& 4$\times$4& 2& 1& 256 & 64 &concat with layer \#2\\
14 & deConv2d.& 4$\times$4& 2& 1& 128 & 3 &concat with layer \#1\\
\hline
\end{tabular}
\end{table*}

\begin{table*}[thb]
\centering
\caption{Architecture of both Reference Reveal network and Residual Reveal network. Each layer is a Inception block with two kinds of kerneal size(3$\times$3 and 5$\times$5), and  the output is the concatenation of both feature maps .There is a batch normalization layer(BN) and a Rectified Linear Unit(ReLU)  after each block layer except the last one.The output convolution layer is followed by a $Sigmoid$ function.}\label{tab:r}
\begin{tabular}{ccccccc}
\hline
Index & Type& Kernel& Stride&Padding& Input & Out\\
\hline
1 & Conv2d.& 3$\times$3 and 5$\times$5 & 1 & 1 and 2& 3 & 50$\times$2\\
2 & Conv2d.& 3$\times$3 and 5$\times$5 & 1 & 1 and 2& 100 & 50$\times$2\\
3 & Conv2d.& 3$\times$3 and 5$\times$5 & 1 & 1 and 2& 100 & 50$\times$2\\
4 & Conv2d.& 3$\times$3 and 5$\times$5 & 1 & 1 and 2& 100 & 50$\times$2\\
5 & Conv2d.& 3$\times$3 and 5$\times$5 & 1 & 1 and 2& 100 & 50$\times$2\\
5 & Conv2d.& 1$\times$1 & 1 & 0& 100 & 3\\
\hline
\end{tabular}
\end{table*}

\vspace{0.05in}
\noindent \textbf{Step-1: Reference/Residual Frame Labeling}: We adopt a simple thresholding approach for labeling a frame to be reference or residual type. Specifically, the first frame in a video is surely labeled as reference. The following frames in the same video sequentially calculate their \emph{averaged pixel-wise discrepancy} (APD)\footnote{ {\footnotesize For two RGB frames, we calculate pixel-wise absolute difference and take the average for R-, G-, and B-channel respectively. The APD score is defined as the average value across R, G, B channels.}} with respect to the first frame. Once the APD score of any frame exceeds some pre-specified threshold, it will be set as a new reference and used to calibrate all following frames. The procedure proceeds until all frames are labeled.

\vspace{0.05in}
\noindent \textbf{Step-2: Hiding Secret (encoding)}: This step does Alice's job in Fig.~\ref{fig:stego}. The key differentiator of our method to others is a divide-and-conquer scheme. Note that in Fig.~\ref{fig:pipeline} two hiding networks are devised, referred to as \emph{Reference H-net} or \emph{Residual H-net} respectively. Each frame is fed into the corresponding H-net by their label. It should be clarified that these two H-nets do not share any parameter. They are individually optimized for encoding specific type of frame only. In this paper we term the new frame, which appears similar to the cover yet conceals a secret somewhere, as \emph{container}. In practice, we choose the U-net model~\cite{DBLP:journals/corr/ChenPK0Y16,IsolaZZE17} for both H-nets. The network specifications are found in Table~\ref{tab:h}.

\vspace{0.05in}
\noindent \textbf{Step-3: Revealing Secret (decoding)}: It does Bob's job in Fig.~\ref{fig:stego}. The input is merely the container, and the output (we call it \emph{decoded secret}) is another image which is desired to be exactly the secret in the perfect case. Otherwise the model is trained to minimize the discrepancy between the secrete and its decoded version. Similar to H-nets, two R-nets (\emph{Reference R-net} or \emph{Residual R-net}) are introduced to reveal the frame or residual secret. However, unlike the encoding stage, Bob strictly has no access to the cover or secret, which implies that frame labels are missing. State differently, the decoder is not aware of which R-net is the optimal handler. We postpone this decision to the next step. The container frame will be sent to both R-nets and obtained two decoded secret images. The specification of R-nets is found in Table~\ref{tab:r}. It is clarified that two R-nets do not share parameters, despite the same network architecture.

\vspace{0.05in}
\noindent \textbf{Step-4: Frame-or-Residual Classification}: Our proposed temporal residual modeling raises new challenges to the classic scheme as depicted In Fig.~\ref{fig:stego} - Bob receives two copies of decoded secret messages in Step-3, from Reference R-net or Residual R-net respectively. Clearly, only one of the secrete message is true. Bob needs to pick out the real message. In fact, we can exhaustively enumerate all possible messages: the real reference and fake residual (container with a true reference secret gets through Reference and Residual R-nets respectively), real residual or fake reference (similar to above, but containers now carry residuals), totalling four valid cases. Therefore, we formulate it as a four-way classification problem. As seen in Fig.~\ref{fig:pipeline}, a Reference-or-Residual (RoR) Net is devised for judging an input decoded message.

Similar to the R-nets, we let RoR Net have a mainframe of five convolutional layers, each of which is paired with BN layer and LeakyReLU. The key difference to R-nets is that the network head is a linear fully-connected layer followed by some softmax layer. Given an input image, the softmax eventually returns a 4-d probabilistic vector that categorizes the decoded information.

\vspace{0.05in}
\noindent \textbf{Step-5: Residual Frame Reconstruction}: This step is optional if Step 4 judges a message as real reference. However, for a residual frame, it is not visually understandable per se. One need to add decoded residuals to the correct reference frame for obtaining the concealed video frame. Since we always process video frames in temporal order, we can record the latest reference frame for reconstructing residuals.

In our proposed system, H-nets / R-nets are jointly trained before the RoR net. The overall loss function for learning H-nets / R-nets is composed as individual loss defined on each networks. Recall that H-nets output container frame and R-nets return decoded reference or residuals. Following a typical treatment in image segmentation, for H-nets we define a loss on H-nets as summing all pixel-wise difference between container / cover, and a loss on R-nets for comparing decoded references / residuals and the original copies. For learning the RoR net, we adopt the standard cross-entropy loss to enforce label consistency.

\section{Experiments}
\label{sec:exp}

\subsection{Dataset Description and Experimental Setting}

There is no available benchmark used for video steganography research. We therefore construct a new benchmark as follows: TRECVID Multimedia event detection (MED)\footnote{http://www-nlpir.nist.gov/projects/tv2017/Tasks/med/} is a yearly competition about retrieving specific semantic events (such as ``birthday party" or ``parkour") from a huge pool of videos. The MED 2017 video corpus consists of more than 0.3 Million videos with high-quality annotation. Since our task is essentially unsupervised, we ignore the video semantic labels and randomly sample 12,000 videos from the whole set. For each video, a 2-second clip is randomly cropped and 24 frames are extracted using the tool of FFMPEG. We generate a data split of training / validation / testing subsets, with 10,000, 1000, and 1,000 video clips respectively.

On all 10,000 training videos, our simple thresholding scheme generates 43,610 reference frames and 196,840 residuals. Videos are randomly drawn to form the (cover, secret)-pair. The Reference H-net is trained using all reference frames, and Residual H-net utilizes the residuals. All decoded messages collectively trains the four-way RoR net. We tune the network parameter following common tactics, such as decaying the learning rate after a fixed number of iterations and using momentum to keep the solution stable. The best model on the validation set is kept as the final model.

\subsection{Empirical Evaluation and Analysis}

\begin{figure*}[t!]
\begin{center}
\includegraphics[width=0.9\linewidth]{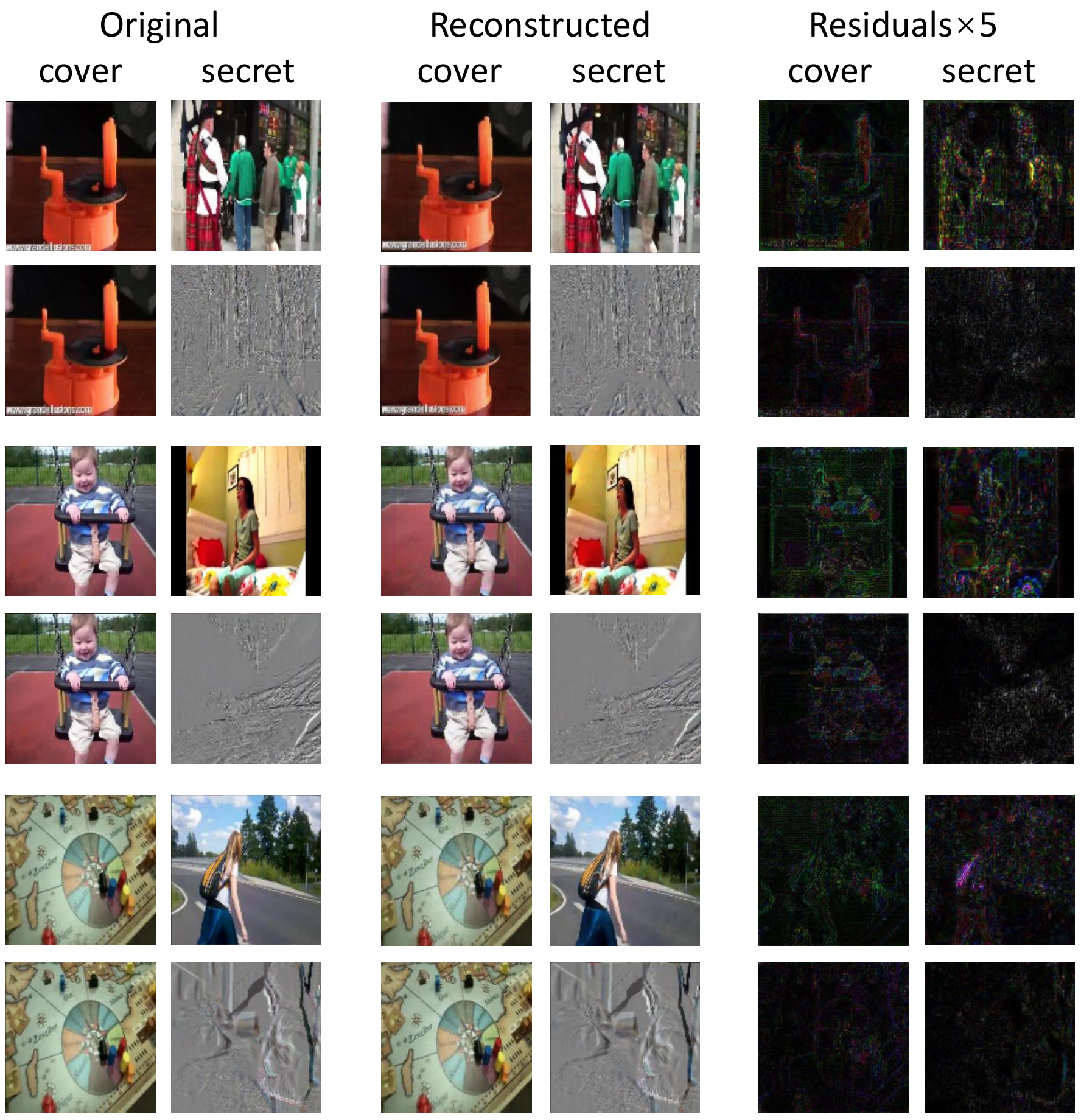}
\end{center}
   \vspace*{-0.1in}
   \caption{\small Hiding results using our video model. Left pair of each set: original cover and secret frame. Center pair: cover frame embedded with the secret frame (container), and the decoded secret frame. Right pair: Residual errors for container and secret (enhanced 5x). Secret frames in odd and even rows are reference frames and residual frames respectively.}
\label{fig:res0}
\end{figure*}

\begin{figure*}[t!]
\begin{center}
\includegraphics[width=\linewidth]{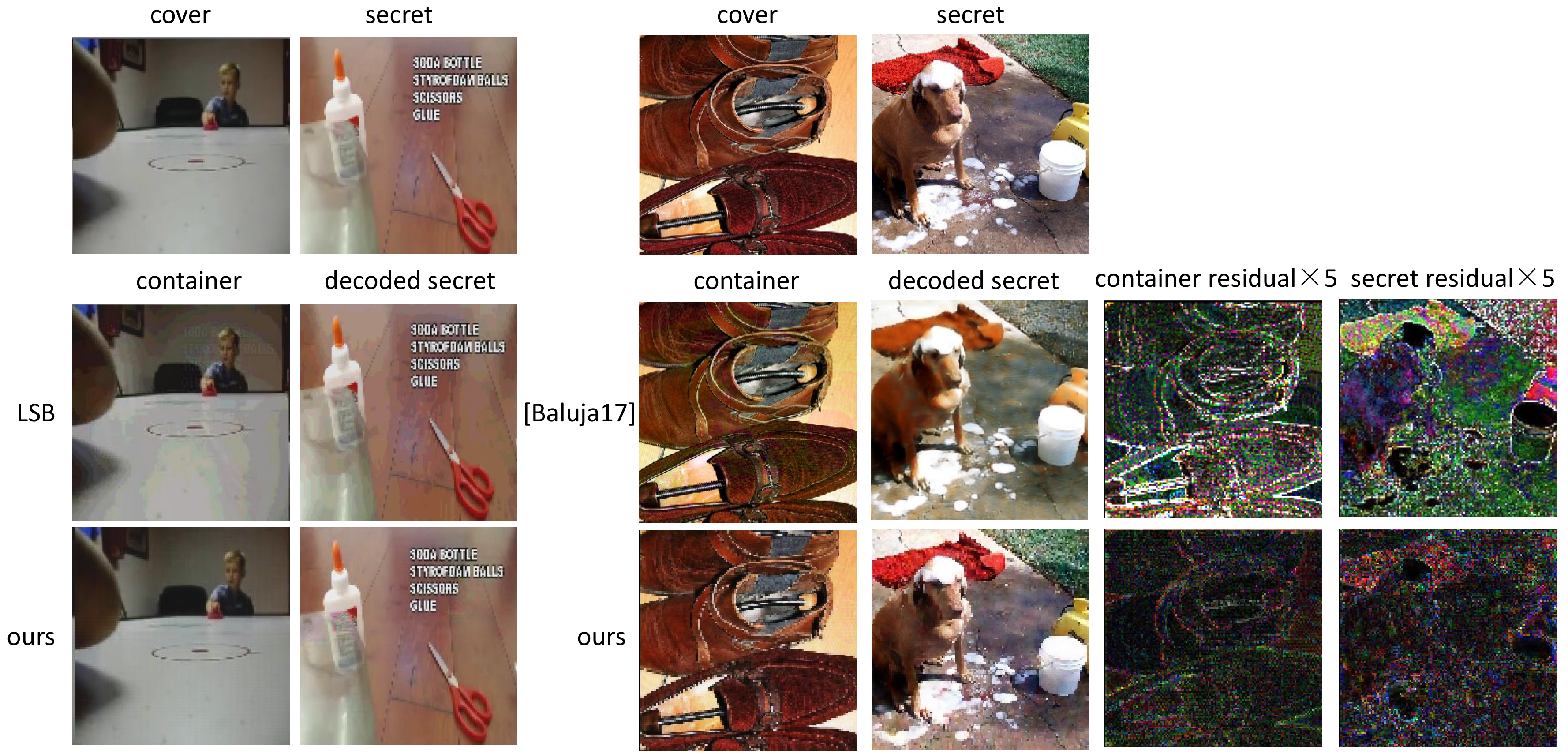}
\end{center}
   \vspace*{-0.1in}
   \caption{\small Comparison with LSB and Deep Steganography~\cite{Baluja17}. Left pair: The proposed method outperforms LSB in perceptual quality. Right pair: Our model achieves better color fidelity and minor residual error than image model.}
\label{fig:comparison}
\end{figure*}

Fig.~\ref{fig:res0} shows the steganography results on selected videos. For each video, we show both the results of Reference H/R-nets and Residual H/R-nets. It is observed that secret videos often have different color tone and textures from the covering video, constructing a challenging task. By investigating the residuals between container-cover and secret-decoded secret pairs as in Fig.~\ref{fig:res0}, one can observe that the container frames still look visually natural. In Table~\ref{table:comparison}, we report the APD scores for the baseline least significant bits (LSB). Since our Reference H/R-net is alike to the work in~\cite{Baluja17}, we separately report the APD scores of Reference H/R-nets (\emph{image model} in the table) and the full architecture in Fig.~\ref{fig:pipeline}. It is seen that our full model enjoys few distortions for the container frames. This may attribute to the separate processing of reference / residual frames. The decoded secret message by the full model is slightly worse, potentially caused by the residual reconstruction step. Both the image model and our video-based model significantly outperforms LSB. We also perform visual comparison with LSB and~\cite{Baluja17}\footnote{The results are based on an unofficial re-implementation of this image steganography model in https://github.com/harveyslash/Deep-Steganography.} in Fig.~\ref{fig:comparison} and clear superiority goes to our model.

\begin{table}[tbp]
\centering
\begin{small}
\begin{tabular}{lcc}
&  $\|$ container - cover $\|$  &  $\|$ secret - decoded secret $\|$ \\
LSB      model &6.64 &8.64 \\
Image model&4.46 &5.04 \\
Our      model&3.80 &5.84\\
\end{tabular}
\end{small}
\caption{Averaged pixel-wise discrepancy (APD) scores for different methods.}
\label{table:comparison}
\end{table}

\subsubsection{Where is the Secret Frame Encoded?}

\begin{figure*}[t!]
\begin{center}
\includegraphics[width=0.8\linewidth]{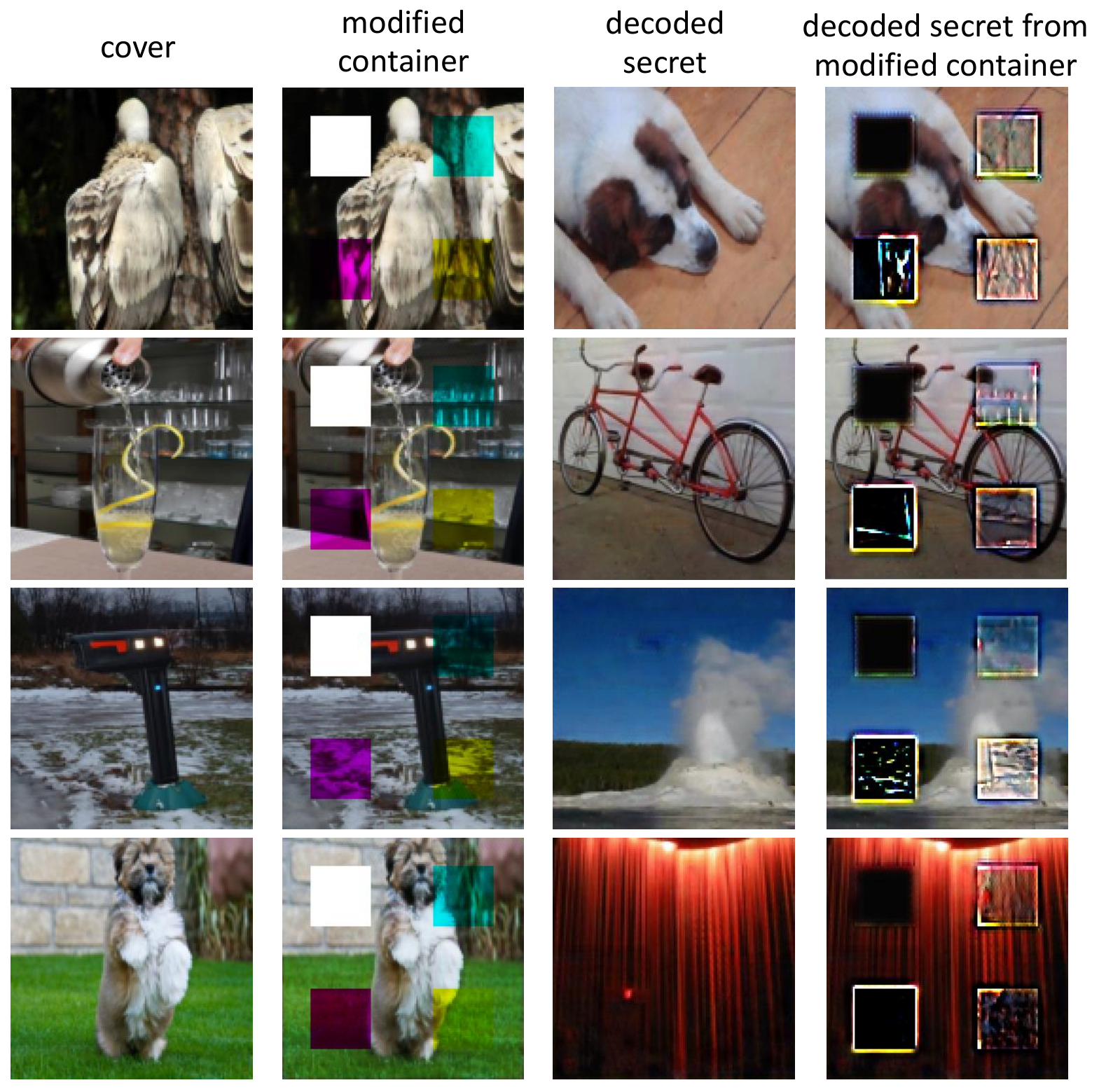}
\end{center}
   \caption{\small Finding how the secret message is concealed. Column 1: cover image. Column 2: modified container by manually adjusting four patches of original container image. Column 3: original decoded secret. Column 4: decoded secret image by re-feeding modified container to R-net.}
\label{fig:encoding_location}
\end{figure*}

Following~\cite{Baluja17}, we manually adjust the pixel values in containers and re-feed the modified container to R-net. By checking the changes in decoded secret frame one may discover where the secret is hidden. As illustrated in Fig.~\ref{fig:encoding_location}, we adjusted four 16$\times$16 areas in a container. The left top one sets all pixels to white color, and other three set either of the RGB channels to 0, respectively. We have several observations: first, the decoded secret changes accordingly, with a large spatial field. It implies our encoding / decoding models are non-local, unlike LSB. Secondly, we find that image gradient matters in visual steganography. Zeroing a single color channel does not change much the image gradient of the inner pixels, which explains that the inner region is still accurately reconstructed. However, the boundary pixels, which suffers from severe gradient change, lose the correct secret message.

\subsubsection{Investigation on Adversarial Learning}

\begin{figure*}[t!]
\begin{center}
\includegraphics[width=0.55\linewidth]{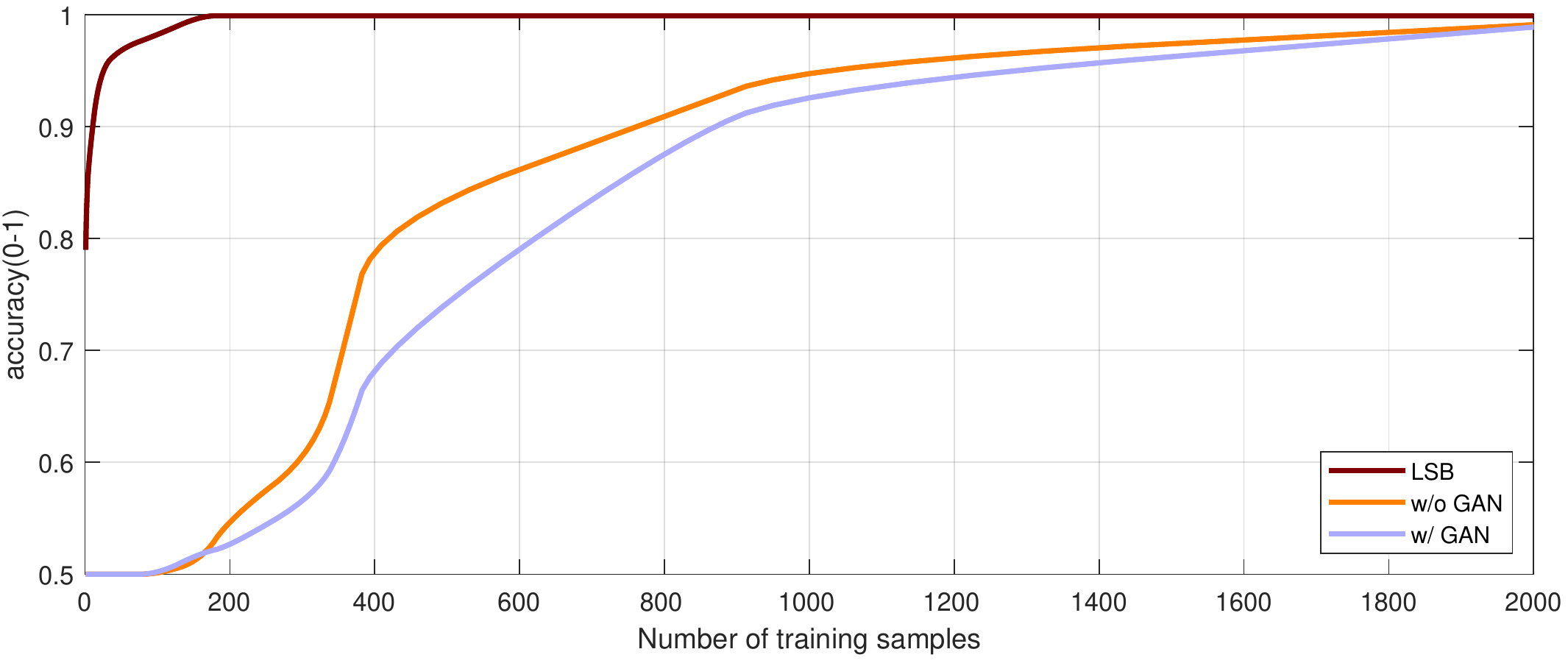}
\end{center}
   \caption{\small Investigation on adversarial learning. The curves indicate how many leaked training samples are needed for an adversary to distinguish a container from a cover, which reflects the anti-steganalysis ability of a model.}
\label{fig:gan}
\end{figure*}

In steganography, a prominent goal is to fool the adversary, Eve in Fig.~\ref{fig:stego}. An interesting problem to us is: after collecting how many labeled cover / container data, the adversary will become accurate enough to detect the presence of secret message? Without loss of generality, we assume the adversary uses a 6-layer CNN for learning a binary classification from labeled data. We investigate both LSB and our video model, as shown in Fig.~\ref{fig:gan}. Interestingly, both methods tend to have zero probability of fooling the adversary after about 2,000 labeled data are leaked.

To increase the resistance to the adversary, we explore an idea of adding an adversarial learning sub-model, similar to~\cite{hayes17}. Specifically, besides H/R-nets we incorporate an adversarial discriminator (we assume it has a common CNN architecture). It can receive frames from the cover and container to make a judgment. If the discriminator cannot distinguish a cover or container, it means that the container generated by H-nets is able to fool this discriminator. We adopt the classic GAN~\cite{GoodfellowPMXWOCB14} loss for this discriminator. In one of our experiments, we combine this GAN loss with H/R losses, modulated via a hyper-parameter (set to 3e-4 empirically), and re-train the video model. The APD scores of the re-trained H/R-nets are 4.13 and 6.21, respectively, which are slightly worse than the original video model in Table~\ref{table:comparison}. This implies that the new loss sacrifices APD in H/R-nets for lower detectability by the adversary.

After re-training, we plot the curve of the new model in Fig.~\ref{fig:gan} (denoted as ``w/ GAN"). The apparent gap between the curves of original model (denoted as ``w/o GAN") and new model indicates that the latter is enhanced for anti-steganalysis. For example, for the original model, leaking 400 training pairs can enable the attacker to correctly distinguish 80\% testing samples. While for the adversarially-trained new model, to achieve this accuracy, more than 600 pairs are required. This experiment serves a strong evidence that incorporating a GAN-style adversarial discriminator can lead to a more steganalysis-secure message embedding. It is also noted that, for LSB the adversary can easily perform shift operations on covers and containers to distinguish, making it less secure.

\subsubsection{The Good and Bad Cover / Secret}

\begin{figure*}[t!]
\begin{center}
\includegraphics[width=0.9\linewidth]{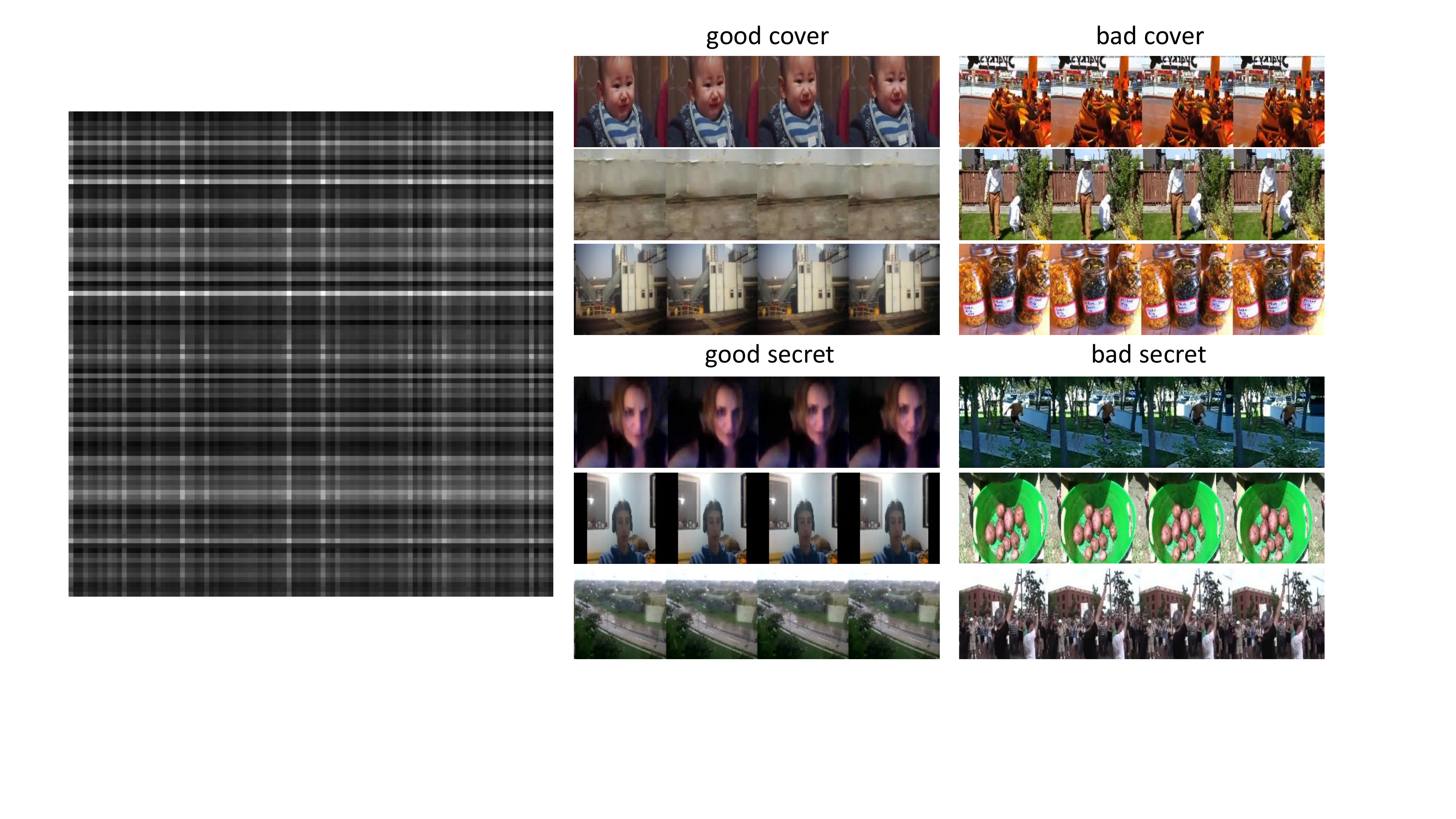}
\end{center}
   \vspace*{-0.7in}
   \caption{\small Investigation of the ``goodness" of covers or secret videos. Left: the matrix of inter-video APD scores (darker ones indicate lower APD scores). Right: examples of ``good" or ``bad" covers or secret.}
\label{fig:good_pair}
\end{figure*}

It is an import problem to predict whether a pair of cover / secret has the potential to generate low APD score. To this end, we randomly select 100 videos as covers, and other 100 videos as secret videos to be hidden. They are exhaustively paired and tested by our video model. We record the pairwise APD scores and plot them on the left matrix in Fig.~\ref{fig:good_pair}. Interesting, we find that some videos tend to be a ``good" cover or secret whatever the other party in the pair is, and vice verse. It becomes clear by looking at the white or black stripes in Fig.~\ref{fig:good_pair}. For better understanding, we select the top 3 best or worse cover / secret and show them in the right of Fig.~\ref{fig:good_pair}. It is observed that ``good" ones tend to have few textures and lower saturation.

\begin{figure*}[t]
\begin{center}
\includegraphics[width=0.42\linewidth]{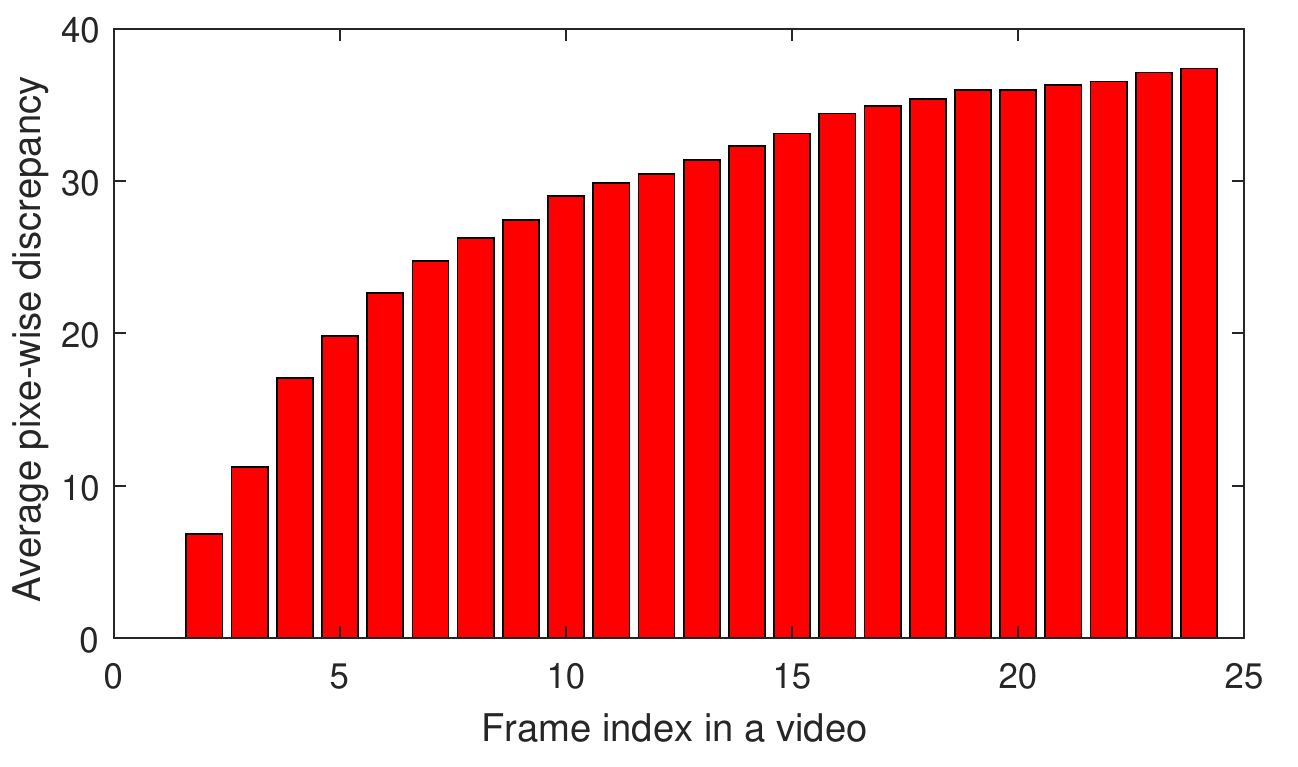}
\includegraphics[width=0.44\linewidth]{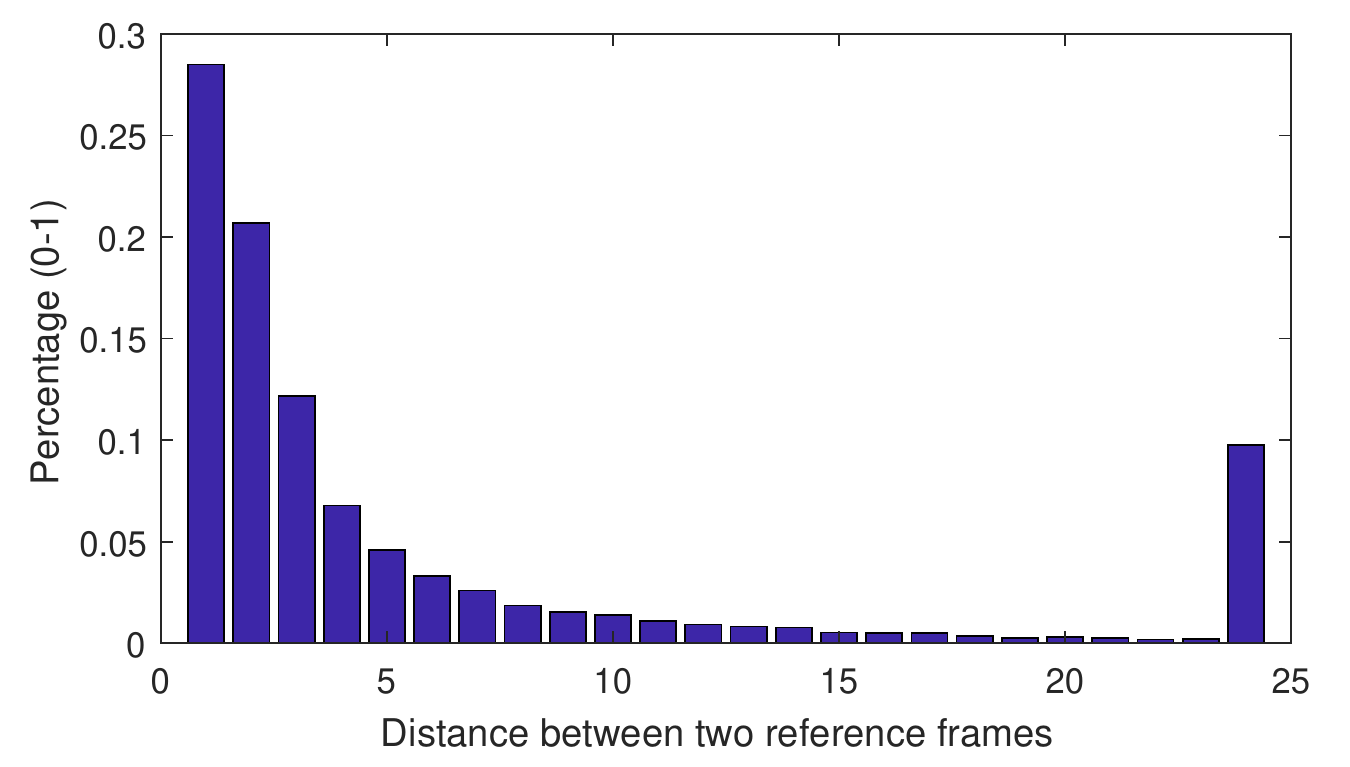}
\end{center}
   \caption{\small Left: APD score between all 24 frames in a video with respect to the first. Right: the distribution of the lengths of video segments obtained via thresholding.}
\label{fig:loc}
\end{figure*}

\subsubsection{Effect of Frame Locations}

We adopt a thresholding scheme to split reference and residual frames. However, choosing a proper threshold is non-trivial. In our experiment, we randomly selected 1000 short videos (24 frames each one) and calculated the average of every pixel's residual from each frame to the first frame. The curve is showed in Fig.~\ref{fig:loc}. A large threshold will generate more residuals, which tends to lead improved container quality yet may degrade the decoded secret. If we set a smaller threshold, there will be more reference frames, making the video model quickly converge to the image steganography. To search a good balance between H/R-nets, we finally selected a threshold of 30.68. Let us term a reference frame and its following residuals as a \emph{video segment}. The right one in Fig.~\ref{fig:loc} plots the distribution of video segment lengths under the chosen threshold.

\begin{figure*}[t]
\begin{center}
   \includegraphics[width=0.85\linewidth]{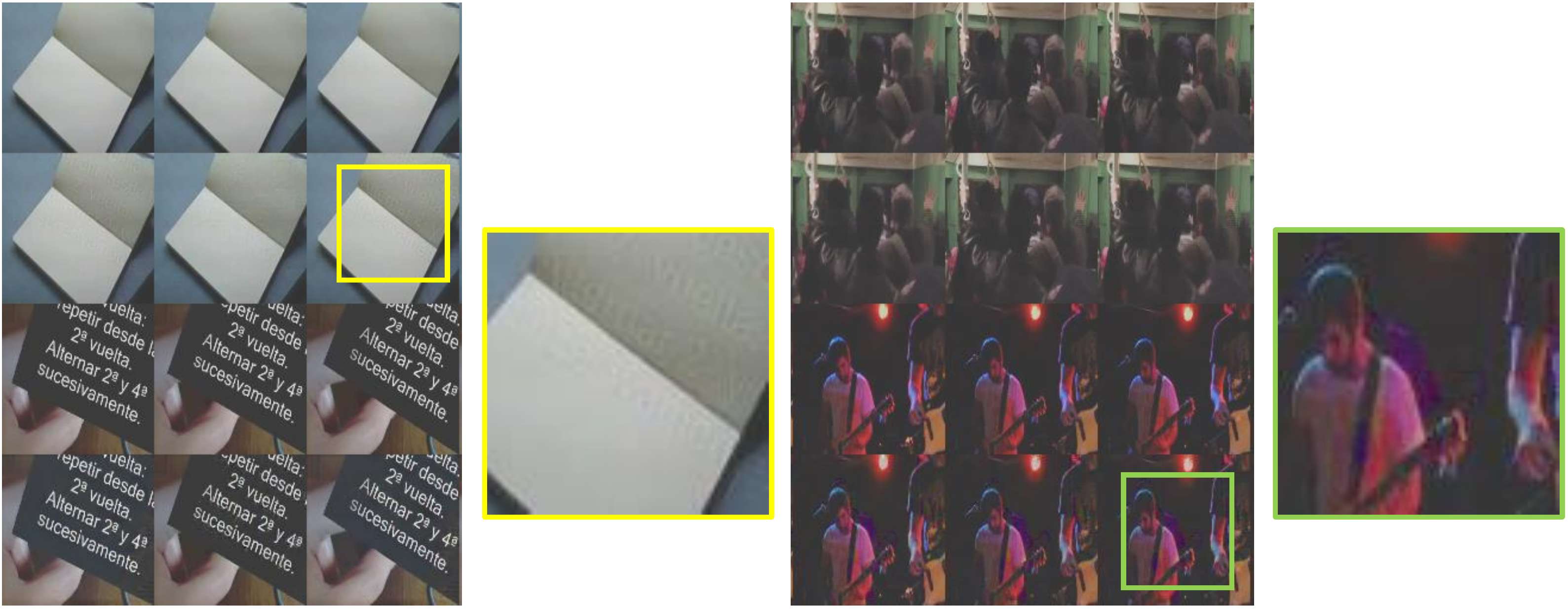}
\end{center}
   \caption{\small Failure cases.}
\label{fig:failure}
\end{figure*}

\subsubsection{Reference-or-Residual (RoR) Network}

As stated earlier, to categorize the decoded message we train a four-class CNN Reference-or-Residual (RoR) classifier. In practice, we use the trained Reference H/R-nets and Residual H/R-nets to collect training data, and use these data to train the RoR network. On the testing set, an accuracy of 99.9625\% was achieved, which is nearly perfect yet the RoR network is still fooled by some hard samples. To attack this issue, we propose an improved judgment method. Because there are only two combinations of reference and residual values, i.e. real reference and fake residual or fake reference and real residual. Therefore, we add the probability of real reference and probability of fake residual as P1, the probability of fake reference and the probability of real residual was added up as P2. If P1 is larger than P2, we suppose that this container conceals reference information, otherwise it hides residuals. This simple scheme brings a 100\% accuracy on the test set.

\section{Concluding Remarks}
\label{sec:conclusion}

In this paper we present a novel deep neural network for the task of video steganography. To fully utilize the sparse property of inter-frame differences, we develop a temporal residual modeling technique, separately treating reference and residual frames during generating steganographic videos. Comprehensive evaluations and studies show the superiority of our method. The future work shall include the exploration of more sophisticated deep models, such as C3D~\cite{DBLP:journals/corr/TranBFTP14}, which may better handle failure cases as in Fig.~\ref{fig:failure}.

{\small
\bibliographystyle{ieee}
\bibliography{egbib}
}

\end{document}